\documentclass[%
 reprint,
nofootinbib,
showkeys,
 amsmath,amssymb,
aps,
]{revtex4-2}

\usepackage{graphicx,xcolor}
\usepackage{dcolumn}
\usepackage{bm}
\usepackage{braket,physics}
\usepackage{mathtools}

\usepackage{hyperref}

\begin{document}

\preprint{APS/123-QED}

\title{Tunnelling theory of Weyl semimetals in proximity to a metallic band}

\author{L. Goutte}
  \email{leo.goutte@mail.mcgill.ca}
\author{T. Pereg-Barnea}%
\affiliation{%
 Department of Physics and Centre for the Physics of Materials \\
McGill University, Montréal, Québec, H3A 2T8, Canada
}%

\date{\today}

\begin{abstract}
We study the effects of tunnelling on the band structure and Fermi arc of a time-reversal broken Weyl semimetal (WSM). 
When coupled to a non-magnetic parabolic band, the WSM's chiral arc state lowers in energy and forms, together with a previously extended state, a noticeable spin-dependent asymmetry in the interface spectrum in the vicinity of the Weyl nodes. 
We study these effects with a lattice model which we solve numerically on a finite sample and analytically through using an ansatz on an infinite sample.  Our ansatz agrees very well with the  numerical simulation as it accurately describes the behaviour of the chiral state, from its energy asymmetry to the spin canting at the interface.
We find that the tunnelling effectively increases the Fermi arc length, allowing for the presence of interface states beyond the bare Weyl nodes.  These additional states may carry current along the interface and their contribution can be detected in the conductance. 
Across the interface, the spin-independent conductance reproduces the results of an electron tunnelling experiment to reveal the WSM's density of states. 
Besides conductivity, the effect of tunneling between the WSM and the metallic band can be seen in quantum oscillations experiments which we briefly comment about.
\end{abstract}

\maketitle


\section{Introduction}
Weyl semimetals (WSMs) are materials whose low-energy excitations are Weyl fermions \cite{RevModPhys.90.015001,Shen2012,weylcoming}. While these particles have their roots in high-energy physics as solutions to the massless three-dimensional Dirac equation in a chiral basis, WSMs present an elegant way of accessing their properties in the condensed matter regime.
A growing interest in these materials culminated with their physical realization in TaAs \cite{taasweylexperiment} and TaNb\cite{tanbweylexperiment}, with additional predictions of type-II WSMs in $\text{WTe}_2$ \cite{typeiiwsmwte2} and $\text{MoTe}_2$ \cite{type11wsmmote2}. On the theoretical side, the WSM's classification as a gapless topological phase makes it an appealing object of study with deep connections to topological Chern insulators \cite{reviewoftopologicalphasesthinfilm} and novel properties in the presence of superconductivity \cite{weylsuperconductor,weylmajoranaflatband} and external magnetic fields \cite{burkovweyl}, to name but a few. 

The Weyl Hamiltonian describes a linear crossing of two non-degenerate bands. For a pair of such bands to touch, one must in general tune three independent parameters, one for each Pauli matrix. In three spatial dimensions with three independent momenta Weyl points are therefore robust against weak perturbations. Near these points/nodes the bulk energy disperses linearly and the physics are governed by the Weyl Hamiltonian:
\begin{equation}
    H = \hbar \mathbf{v}_0 \cdot \mathbf{k} \pm \hbar v \mathbf{k} \cdot \bm{\sigma},
\end{equation}
where $\pm$ denotes the node's chirality, $v$ is the effective Fermi velocity, $\mathbf{k}$ is the momentum and $\bm{\sigma}$ is the vector of Pauli matrices acting in spin space. The first term, proportional to the unit matrix, breaks Lorentz invariance and tilts the dispersion. For type-I WSMs, it can be ignored, leaving only the second term. The latter has a linear dispersion that, while strongly reminiscent of two-dimensional graphene, will not open a gap in the presence of small perturbations. Each Weyl node is also a monopole of Berry curvature, leading to a chiral anomaly which manifests itself in many exotic properties such as the Quantum anomalous Hall effect, negative magnetoresistance \cite{negativemagnetoresistance}, the chiral magnetic effect \cite{obrienchiralmagneticeffect}, and high carrier mobility \cite{highmobility}.

In lattice systems, a Weyl semimetal hosts pairs of Weyl nodes along a given nodal direction \cite{Burkovnodalsemimetals,burkovwsmmultilayer,RevModPhys.90.015001}. This is required by either time-reversal ($\mathcal{T}$) or inversion ($\mathcal{I}$) symmetry and the fact that the total Berry flux in the first Brillouin zone (BZ) must vanish. One can slice the system along the nodal direction and assign a Chern number to each two dimensional slice of momentum space: if a plane is pierced by Berry flux it will be topologically non-trivial, and vice-versa. Therefore, the bulk-boundary correspondence implies the presence of topologically protected surface states in between the Weyl nodes only. At the Fermi level, then, an open system will host a \textit{Fermi arc} -- a projection of zero-energy chiral surface states connecting pairs of opposite chirality Weyl nodes and dispersing linearly away from the Fermi level. In this sense, gapless topological phases are intermediaries between genuine trivial and topological phases of matter and can even be realized by a repeated stacking of the two \cite{burkovwsmmultilayer}. 

While a growing number of their properties are known, such as the effect of impurities and defects \cite{impuritiesintypeii,impuritiesintypeii2,Silva_2022,rkkyinteractioninwsms}, the manoeuvrability and theoretical richness of these materials further motivates the analytical study of tunnelling in WSMs. In what follows, we investigate the effect of tunnelling by constructing a tight-binding model of a time reversal symmetry ($\mathcal{T}$) broken WSM coupled to a non-magnetic band. In describing a single Fermi arc, the $\mathcal{T}$-broken WSM displays all aforementioned properties while providing a minimal model to serve as a building block for setups with more pairs of nodes. Likewise, our choice of a simple tunnelling potential and featureless band are intentional: we seek to draw out the bare properties of a WSM in contact with a non-topological material. 

The remaining sections are structured as follows. In Sec.~\ref{sec:model}, we present the WSM and non-magnetic band models along with the specific form of surface tunnelling. The numerical results of a finite lattice model are then presented in Sec.~\ref{sec:finitelatticemodel}. In Sec.~\ref{sec:discretetheory} we derive an infinite lattice theory with an interface to model the spectra, spin canting and interface arcs in a lattice framework, while Sec.~\ref{sec:interfacetheory} presents a simpler continuum model. We finish in Sec.~\ref{sec:transport} by investigating the novel transport properties of the coupled system both along and across the interface in the Landauer-Buttiker and electron tunnelling formalism, respectively. Directions for further study are briefly touched upon in the conclusion, Sec.~\ref{sec:conclusion}, and relevant technical details are included in the appendices.

\section{Model}
\label{sec:model}

\subsection{Weyl semimetal}
We consider a minimal Hamiltonian which captures the Fermi arc feature. This can be achieved either by breaking $\mathcal{T}$ while preserving $\mathcal{I}$ or vice-versa. In order to work with smaller matrices, we choose the former. Explicitly, then, our Hamiltonian must satisfy $ H\left(\mathbf{k}\right) = \sigma_z  H\left(-\mathbf{k}\right) \sigma_z$ and $H\left(\mathbf{k}\right) \neq \sigma_y  H^*\left(-\mathbf{k}\right) \sigma_y $. A simple tight-binding Hamiltonian which abides by these symmetries is ($\hbar = \text{lattice constant} = 1$) \cite{RevModPhys.90.015001}
\begin{subequations}
\label{eq:wsmhamiltonian}
\begin{gather}
    {H}_w = \sum_{\mathbf{k}} \mathbf{c}^{\dagger}_{\mathbf{k}} \mathcal{H}^{\mathrm{bulk}}_w \left(\mathbf{k}\right) \mathbf{c}_{\mathbf{k}}, \\
     \mathcal{H}^{\mathrm{bulk}}_w \left(\mathbf{k}\right) = t_x \sin{k_x}  \sigma_x + t_y  \sin{k_y} \sigma_y +t_z  m\left(\mathbf{k}\right) \sigma_z, \\
     m\left(\mathbf{k}\right) = \left(2 + \gamma - \cos{k_x} - \cos{k_y} - \cos{k_z}\right).
\end{gather}
\end{subequations}
Here, $\mathbf{c}_{\mathbf{k}} = \left(c_{\mathbf{k},\uparrow}, c_{\mathbf{k},\downarrow}\right)^{\top}$ is an annihilation operator in momentum space, $t_{s}$ ($s=x,y,z$) is the strength of hopping in the $s$-direction and ${\sigma}$ are the Pauli spin matrices. We further set $t_x=t_y=t_z=t>0$ for simplicity. 
The Hamiltonian \eqref{eq:wsmhamiltonian} admits the bulk energies 
\begin{equation}
    E_{\pm} = \pm t \left[\sin^2{k_x}+\sin^2{k_y} + m^2\left(\mathbf{k}\right) \right]^{\frac{1}{2}}.
\end{equation}
These vanish at $\mathbf{k}^{\pm}_w = \left(0,0,\pm \arccos{\gamma}\right) \equiv \left(0,0,\pm k_w\right)$ -- the aforementioned Weyl nodes. We emphasize the importance of the $\cos{k_{x,y}}$ terms, without which there would be more than two nodes in the BZ for a given $\gamma$. 

\begin{figure}
    \centering
    \includegraphics[width = \columnwidth]{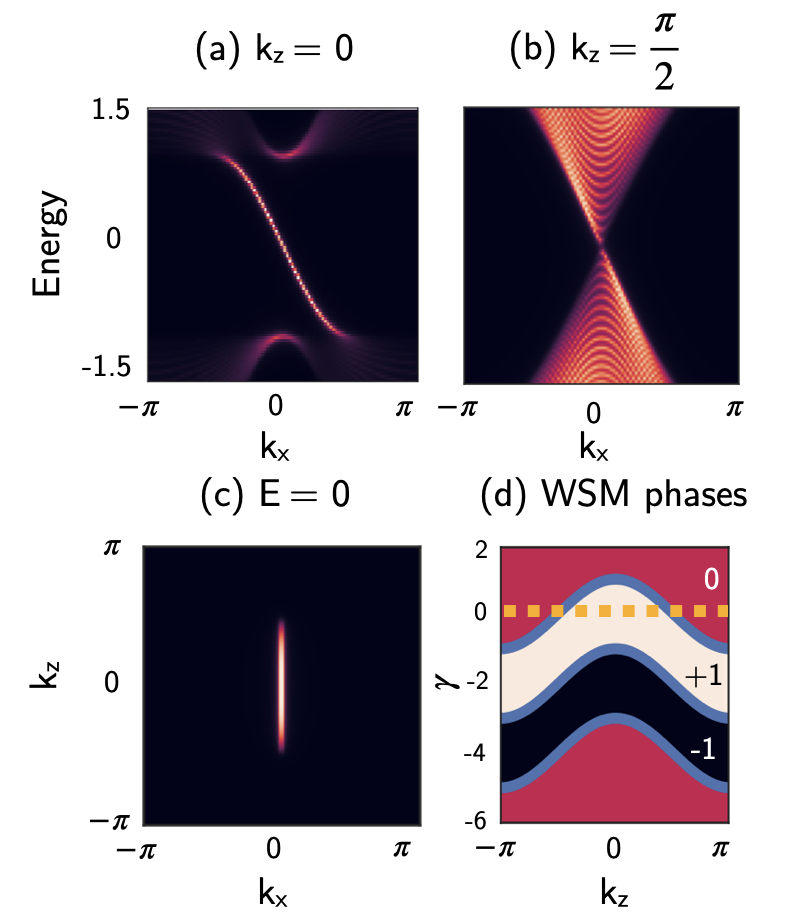}
    \caption{The minimal Weyl semimetal model. (a) Spectral function at $y=L_y-1$ of a WSM open in $y$ plotted in the $E$-$k_x$ plane for fixed  $k_z = 0$ and (b) $k_z = \pi / 2$. (c) WSM spectral function plotted in the $k_x$-$k_z$ plane for fixed $E = 0$ showing the Fermi arc. (d) Phase diagram of Eq.~\eqref{eq:wsmhamiltonian} with lower band's Chern numbers. The phase boundaries $\gamma = \cos{k_x}$, $\gamma = \cos{k_x} - 2$ and $\gamma = \cos{k_z} - 4$ are plotted in blue. We will work at $\gamma=0$ (dashed orange line). 
    }
    \label{fig:barewsm}
\end{figure}

These gapless bulk momenta $\mathbf{k}^{\pm}_w$ suggest that $H_w$ exhibits different phases that depend solely on the arc length parameter $\gamma$. For $\gamma > 1$, $m\left(\mathbf{k}\right) > 0$ for all $\mathbf{k}$ and the system is trivially gapped. As $\gamma$ decreases to $1$, a pair of Weyl nodes appear at the origin and move outward along $k_z$ as $\gamma$ decreases further. This defines a gapless topological phase whereby a nonzero Berry flux flows within the momentum range $|k_z| < k_w$ from the node of negative chirality to the one of positive chirality. Consequently, the Chern number -- defined for a fixed $k_z$ -- is nonzero between the nodes, and zero beyond them. When $\gamma \leq -1$, the Weyl nodes reach the BZ boundaries and disappear, leaving the bulk dispersion with an inverted band gap. Between $-5<\gamma<-1$, the same process occurs for Weyl nodes with $(k_x,k_y) = (0,\pi)$, $(\pi,0)$, and $(\pi,\pi)$, until $\gamma < -5$ where the system is again gapped and trivial for all $\mathbf{k}$. In all numerical results that follow, we take $\gamma = 0$ ($k_w = \pi / 2$), well within the gapless topological regime and with a Fermi arc length $k_{\mathrm{arc}} = \pi$. The bare WSM's surface spectrum, Fermi arc, and topological phases are shown in Fig.~\ref{fig:barewsm}.

\subsection{Tunnelling}

\begin{figure}
    \centering
    \includegraphics[width = \columnwidth]{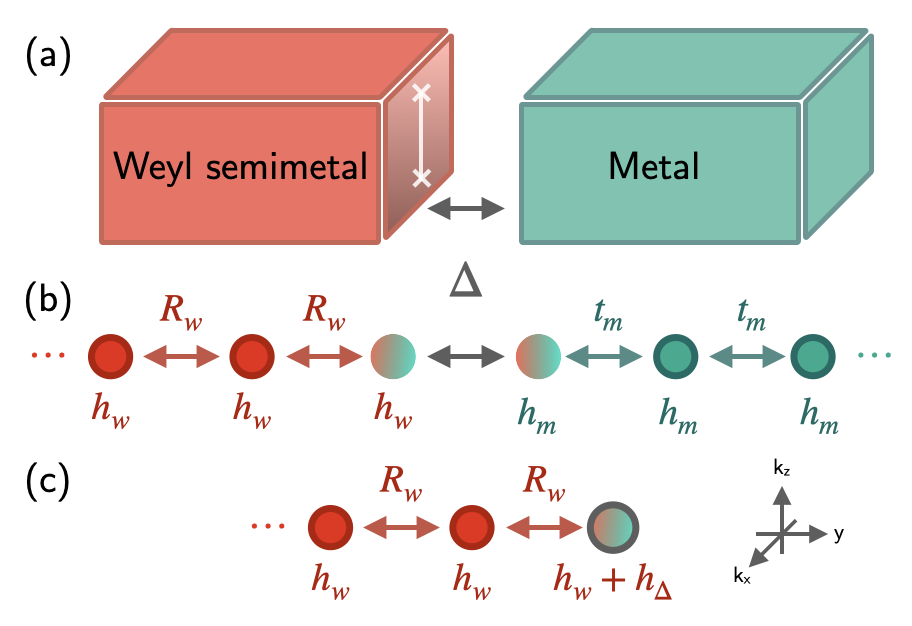}
    \caption{(a) Schematic of the WSM-metal system. Only the rightmost surface, or \textit{interface} (with the Fermi arc shown as a white line and the nodes as white crosses) is linked to the metal via tunnelling $\Delta$. (b) Physical representation of the system as a chain ($\mathbf{R}_w$, $\mathbf{h}_w$, etc. defined in App.~\ref{sec:hamiltonianfullform}). (c) By integrating out the metal degrees of freedom, the chain is simplified into a single semi-infinite chain with a single edge site of energy $h_{\Delta} = T^{\dagger} G_m T$ (shaded with grey line).}
    \label{fig:schematic}
\end{figure}

To draw out the tunnelling properties of the Weyl semimetal, we couple it to a simple parabolic band via non-magnetic surface tunnelling. The band's Hamiltonian is spin-independent and reads
\begin{subequations}
\label{eq:metalhamiltonian}
    \begin{gather}
    H_m = \sum_{\mathbf{k}} \mathbf{d}^{\dagger}_{\mathbf{k}} \mathcal{H}^{\mathrm{bulk}}_m \left(\mathbf{k}\right) \mathbf{d}_{\mathbf{k}},  \\
    \mathcal{H}^{\mathrm{bulk}}_m \left(\mathbf{k}\right) = -2 t_m \left( \cos{k_x} + \cos{k_y} + \cos{k_z}\right) - \mu,
\end{gather}
\end{subequations}
where $t_m$ is the hopping amplitude, $\mu$ the chemical potential and $\mathbf{d}_{\mathbf{k}} = \left(d_{\mathbf{k},\uparrow}, d_{\mathbf{k},\downarrow}\right)^{\top}$ is an annihilation operator in momentum space.
For brevity, we equivalently refer to this non-magnetic parabolic band as ``metal", though one may of course tune $\mu$ to achieve a semi-conductor or an insulator, as discussed in App.~\ref{sec:varymetal}, where we also consider two parabolic bands.

We now introduce a tunnelling Hamiltonian which couples the surface of the WSM to the surface of the metal. We proceed with open boundary conditions in the $y$-direction and keep well-defined momenta perpendicular to the surface, $\mathbf{k}_{\perp} = \left(k_x,k_z \right)$. The WSM (metal) side runs from $y=-L_y + 1$ to $0$ ($y=1$ to $L_y$), defining an interface between the WSM's $y=0$ and metal's $y=1$ sites.
The Hamiltonian for the full (finite-sized) system is therefore 
\begin{subequations}
\label{eq:mainhamiltonian}
\begin{gather}
    {H} = \sum_{\mathbf{k_{\perp}}} \sum_{y,y'=-L_y+1}^{L_y} \mathbf{f}_{\mathbf{k}_{\perp},y}^{\dagger} \mathcal{H}\left(\mathbf{k}_{\perp}\right)_{y,y'} \mathbf{f}_{\mathbf{k}_{\perp},y'}, \\
    \mathcal{H}\left(\mathbf{k}_{\perp}\right) = \begin{pmatrix}
\mathcal{H}^{\mathrm{open}}_w \left(\mathbf{k}_{\perp}\right) & T^{\dagger} \\
T & \mathcal{H}^{\mathrm{open}}_m \left(\mathbf{k}_{\perp}\right)
\end{pmatrix},
\end{gather}
\end{subequations}
where
\begin{equation}
    \mathbf{f}_{\mathbf{k}_{\perp},y} = \begin{cases} 
    \mathbf{c}_{\mathbf{k}_{\perp},y} & -L_y+1 \leq y \leq 0 \\
    \mathbf{d}_{\mathbf{k}_{\perp},y} & 1 \leq y \leq L_y
    \end{cases}
\end{equation}
and $\mathcal{H}^{\mathrm{open}}$ is the partial-in-$y$ Fourier transform of $\mathcal{H}^{\mathrm{bulk}}$.
The full form of Eq.~\eqref{eq:mainhamiltonian} is shown in App.~\ref{sec:hamiltonianfullform}. 
The surface tunnelling term is also non-magnetic and takes the form
$\left(T\right)_{y,y'} = \Delta \delta_{0,L_y-1}$, or
\begin{equation}
\label{eq:tunnelling}
    T = \begin{pmatrix}
    0 & \dots & \Delta \\
    \vdots & \ddots & \vdots \\
    0 & \dots & 0
    \end{pmatrix},
\end{equation}
where, for simplicity, we have assumed that $\Delta$ is a real constant that modulates the tunnelling strength between interface sites $y=0$ and $y=1$. Physically, the tunnelling strength can be modified either by varying the metal bandwidth $t_m$ or changing the interface thickness, as suggested by Fig.~\ref{fig:schematic}. There are therefore two competing energy scales at the WSM's interface: the interlayer hopping $t$ pulling the electron towards the bulk and the tunnelling strength $\Delta$ pulling the electron towards the metal.

Before moving on to the finite lattice simulations, we note that the metal dynamics can be exactly integrated out to make way for a modified WSM propagator \cite{borchmannbarnea,marchandfranz}. More precisely, the effective Green's function becomes
\begin{equation}
\label{eq:effectivegreen}
    G_{\mathrm{eff}} \left(i\omega_n\right) = \left[G_w^{-1}\left(i\omega_n\right) - T^{\dagger} G_m \left(i\omega_n\right)T \right]^{-1}
\end{equation}
where $\omega_n$ is the Matsubara frequency and $G_{w,m} = \left(i\omega_n - \mathcal{H}^{\mathrm{open}}_{w,m}\right)^{-1}$ are the bare Green's functions. Substituting in Eqs.~\eqref{eq:metalhamiltonian} and \eqref{eq:tunnelling} and  yields, after some algebra,
\begin{equation}
\label{eq:effectivesamesitepotential}
     T^{\dagger} G_m \left(i\omega_n\right) T  =  -  \frac{\Delta^2 }{\sqrt{\left(i\omega_n - h_m \right)^2-4t_m^2}} \delta_{y,L_y-1}\delta_{y,y'},
\end{equation}
where $h_m = -2 t_m \left(\cos{k_x} + \cos{k_z}\right) - \mu$. Thus, surface tunnelling simply shifts the same-site hopping of the last site (Fig.~\ref{fig:schematic}c). 

\section{Finite lattice model}
\label{sec:finitelatticemodel}

\begin{figure*}
    \centering
    \includegraphics[width = \textwidth]{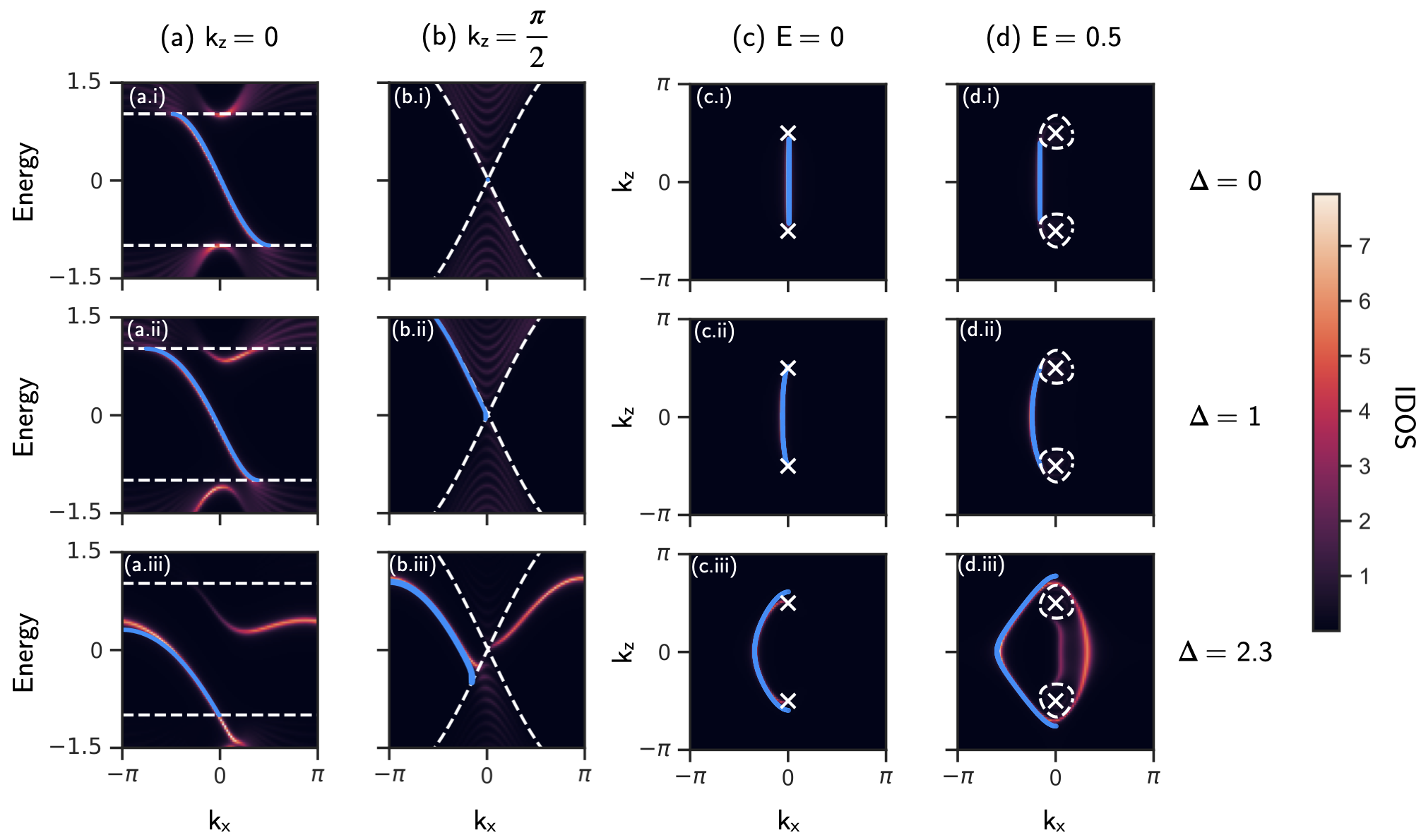}
    \caption{Interface density of states for the coupled WSM-metal system for both spins. The numerical results (simulated on a $L_y = 30$ size chain sampled at $100$ momentum points) are shown in warm colours whereas the chiral state's infinite lattice model  (Sec.~\ref{sec:discretetheory}) is plotted in blue. In all plots where the infinite lattice theory obstructs the numerical results (e.g. the top left), the agreement is near exact. The columns correspond to (a) the spectrum along $k_x$ at $k_z=0$, (b) the spectrum along $k_x$ at the Weyl point $k_z = + \pi / 2$, and the emergent interface arcs at (c) $E = 0$ and (d)  $E=0.5$. The rows are set in increasing order of $\Delta = 0$, $1$, $2.3$ going down. The bulk energy edges $E_{\mathrm{bulk}} = \pm t \left[\sin^2{k_x} + \left(1 + \gamma - \cos{k_x} - \cos{k_z}\right)\right]^{\frac{1}{2}}$ are denoted by dashed white lines, as is the Fermi surface in the $E=0.5$ interface plots. The bare Weyl nodes $\mathbf{k}^{\pm}_{\perp,w} = (0,\pm \pi / 2)$ are white crosses. The fixed parameters used for these and all other plots are $t=1$, $\gamma=0$, $t_m = 0.5$, $\mu = -4$, unless otherwise specified.}
    \label{fig:discreteplots}
\end{figure*}



We now turn to the numerical results of Eq.~\eqref{eq:mainhamiltonian} on a finite lattice. Keeping the system open in $y$ with the quantum numbers $k_x$ and $k_z$, the spectral function is obtained by evaluating $A\left(E,\mathbf{k}_{\perp}\right)  = - {\pi}^{-1} \mathrm{Im} \left[\mathrm{Tr} \left(G\right)\right]$ with the Green's function
\begin{equation}
    G\left(E,\mathbf{k}_{\perp}\right) =\left[{E + i0^{+} - \mathcal{H}\left(\mathbf{k}_{\perp}\right)}\right]^{-1}.
\end{equation}
The WSM's interface density of states (IDOS), displayed in Fig.~\ref{fig:discreteplots}, is found by tracing over the $y = 0$ site only. 

At $k_z=0$ (Fig.~\ref{fig:discreteplots}a), we are exactly in between the Weyl nodes. Without tunnelling, only the so-called \textit{chiral state} is present and localized to the interface, residing on the Fermi arc and dispersing as $E = -t\sin{k_x}$ with a spin $\sigma_x = -1$. With tunnelling, there are two noticeable effects. First, the chiral state lowers its energy as it is now able to hop to the metal side, spreading its wavefunction. Indeed, this lowering of energy captured by Eq.~\eqref{eq:effectivesamesitepotential} is a prevailing effect throughout this work. By that same token, a previously extended state enters the bulk gap from the upper bulk band and localizes to the interface. Contrary to the chiral state, this so-called \textit{emergent interface state} does not have a uniform spin polarization.

At the Weyl nodes (Fig.~\ref{fig:discreteplots}b), the Fermi arc terminates and there are no interface states for $\Delta = 0$. As tunnelling is increased, however, the chiral state can be seen along the Weyl node's upper cone. When $\Delta$ increases beyond the interlayer hopping $t$ the chiral state detaches from the Weyl cone and forms, together with the previously discussed emergent interface state, a noticeable asymmetry in the interface density of states with respect to $k_x$ reflection (Fig.~\ref{fig:discreteplots}b.iii). This striking asymmetry is of particular interest. Physically, it suggests that tunnelling modifies the group velocity along the interface to produce additional left- and right-flowing current in an energy range between the chiral and emergent interface states' intersections with the bulk dispersion. Naively, this is surprising because one may not expect the breaking of translation symmetry in $y$ to induce an asymmetry in the $x$-direction. However, one must remember that the physics on a single surface are not in fact symmetric in $k_x$ to begin with, as evidenced by the linearly dispersing chiral state at the interface. Therefore, although the spectral function is symmetric in $k_x$ when traced over all sites, the localized tunnelling term in $y$ will explicitly break this symmetry. 

By plotting $A\left(E,\mathbf{k}_{\perp}\right)$ in the surface BZ, we see that the zero-energy interface Fermi arc (Fig.~\ref{fig:discreteplots}c and d) will curve in the presence of tunnelling \cite{gorbararcs}. While still terminating at the Weyl nodes $\mathbf{k}^{\pm}_{\perp,w} = (0,\pm k_w)$, it does go beyond $k_z = \pm k_w$ at zero energy, signifying the existence of interface states in a region of parameters outside the bare Fermi arc. This is illustrated by the previously discussed chiral state's presence at $k_z = \pi / 2$ and will have important transport consequences come Sec.~\ref{sec:transportalong}.

These results are robust to changes in the metal's form. In fact, we find that equivalent behaviour may be obtained simply by coupling the WSM to a constant energy reservoir $t_m = 0$, $\mu = -M$. A more realistic setup in which the WSM is coupled to a two-band bulk insulator will yield two copies of the dispersions found in Fig.~\ref{fig:discreteplots}, one for positive and one for negative energy (see App.~\ref{sec:varymetal}). 

As seen in the numerics above, a new closed orbit of low energy states appears on the interface. This closed orbit should be apparent in quantum oscillations experiments as it leads to oscillations with frequency which matches the enclosed momentum space area. These oscillations should be contrasted to the arc/node oscillations suggested by Ref.~\cite{potteroscillations} and studied in Ref.~\cite{borchmannbarnea}. The latter oscillations result from closed orbits which include both the surface (interface) and bulk states, meaning their frequency depends on the slab depth. By contrast, the new orbit seen in Fig.~\ref{fig:discreteplots}d.iii contains only interface states and its frequency is depth independent.


\section{Infinite lattice theory with an interface}
\label{sec:discretetheory}

The physics at the interface seen in the lattice model above can be described in an infinite model and treated analytically with the help of an ansatz. We take $L_y \rightarrow \infty$ and impose $\psi\rightarrow 0$ at $y\to \pm\infty$ on both sides of the interface. Therefore, this theory effectively consists of two semi-infinite slabs connected by surface tunnelling $\Delta$. 

Seeking states $\bm{\varphi}$ exponentially localized to the interface, we make the ansatz
\begin{equation}
\label{eq:wsmdiscreteansatz}
    \bm{\varphi} = \begin{cases}
        \bm{\varphi}_w (y) = e^{ik_xx + ik_zz} \ell^{y}  \bm{\phi}_w & y=-\infty,\dots,-1, 0\\
            \bm{\varphi}_m (y) = e^{ik_xx + ik_zz}  \ell_m^{-y+1}  \bm{\phi}_m & y=1,2,\dots,\infty
    \end{cases}
\end{equation}
where $\bm{\phi}_{w,m}$ are spinors carrying the overall normalization. Note that this ansatz assumes a constant spin direction and is therefore suitable for the chiral state found above but is not completely general. To simplify our problem slightly, we rotate our states by $\pi/2$ about $y$ axis in spin space. Defining $g_1 \equiv t\sin{k_x}$ and $g_3 \equiv t \left(2 + \gamma - \cos{k_x} - \cos{k_z}\right)$ leads to the change
\begin{subequations}
\begin{align}
    \mathbf{h}_w &= g_1 \sigma_z - g_3\sigma_x, \\
    \mathbf{R}_w &= t(\sigma_x + i \sigma_y)/2,
\end{align}
\end{subequations}
for the same-site and nearest-neighbour hopping matrices, respectively. The metal and tunnelling components are unchanged. 

\subsection{$\Delta = 0$}
As a first test of validity, we take the $\Delta=0$ case and recover the chiral state and Fermi arc of the finite lattice model. 
We do not impose any boundary conditions at the $y$-termination but instead just look for exponentially localized states which solve the bulk difference equations. In a lattice formalism, $\mathcal{H}^{\mathrm{open}}_w \bm{\varphi}_w = E\bm{\varphi}_w$ produces a set of coupled difference equations relating $\bm{\varphi}_w(y)$ to its nearest neighbours $\bm{\varphi}_w(y \pm 1)$ \cite{andreevedgestates}:
\begin{equation}
    E\bm{\varphi}_w(y) = \mathbf{h}_w \bm{\varphi}_w(y) + \mathbf{R}^{\dagger}_w \bm{\varphi}_w(y+1) + \mathbf{R}_w \bm{\varphi}_w(y-1) .
\end{equation}
Plugging in Eq.~\eqref{eq:wsmdiscreteansatz}, we obtain the matrix equation
\label{eq:discretedifferenceequations}
\begin{align}
\label{eq:bulkdifferenceequationWSM}
0 = \left(E - g_1\sigma_z + g_3\sigma_x -t\ell^{-1}\sigma_+ - t\ell\sigma_-\right)\bm{\phi}_w
\end{align}
where $\sigma_{\pm} = \left(\sigma_x \pm i \sigma_y \right)/2$. Setting the determinant of Eq.~\eqref{eq:bulkdifferenceequationWSM} to zero yields the ratio of spins and energy, respectively:
\begin{subequations}
\label{eq:discretebulkwsm}
\begin{gather}
\frac{\phi^{\uparrow}_w}{\phi^{\downarrow}_w} = \frac{E+g_1}{t \ell - g_3} = \frac{t\ell^{-1} - g_3}{E - g_1}, \\
\label{eq:discretebulkenergy}
E = \pm \left[g_1^2 +g_3^2 +t^2 -g_3 t \left(\ell + \ell^{-1}\right)\right]^{\frac{1}{2}}.
\end{gather}
\end{subequations} 
Guided by the previous section we make the assumption $\phi^{\uparrow}_w / \phi^{\downarrow}_w = 0$ (spin in the $-x$ direction) and find a solution with $\ell = t / g_3$ and $E = - g_1$. To satisfy the boundary condition at $-\infty$, we impose $\mathrm{Re}\left(\ell \right) > 1$, or $g_3 < t$. The familiar Fermi arc condition $\gamma < \cos{k_z}$ then follows naturally\footnote{For a bulk state of energy $E$, $\mathrm{Re}\left(\ell \right) > 1$ implies $-E_{\mathrm{bulk}}<E<E_{\mathrm{bulk}}$ where $E_{\mathrm{bulk}}(\mathbf{k}_{\perp}) = \left[g_1^2 + (g_3 - t)^2\right]^{\frac{1}{2}}$ is the bulk edge.}. 
We have therefore recovered the aforementioned chiral state: a uni-directional interface state on the Fermi arc. 

\subsection{$\Delta > 0$}

We now allow for tunnelling at the interface between the $y=0$ and $y=1$ sites. There are then four difference equations, one for each type of site: the Weyl bulk, 
Weyl interface, metal interface, and metal bulk. Substituting in the supposed forms of $\bm{\varphi}_{w,m}(y)$, the difference equations are, respectively:
\begin{subequations}
\begin{align}
\label{eq:differenceequationWSM}
    &0 = \left(
    E - g_1 \sigma_z + g_3 \sigma_x - t\ell\sigma_- -t \ell^{-1}\sigma_+\right) \bm{\phi}_w,\\
\label{eq:differenceequationinterface1}
     &  0 = 
     \left(
    E - g_1 \sigma_z + g_3 \sigma_x - t\ell^{-1}\sigma_+\right) \bm{\phi}_w
     -  
    \Delta \bm{\phi}_m, \\
\label{eq:differenceequationinterface2}
    &  0 =
    \left(E - h_m + t_m \ell_m^{-1}\right) \bm{\phi}_m
     -  
    \Delta \bm{\phi}_w,\\
\label{eq:differenceequationmetal}
    & 0 = 
    \left(E - h_m + t_m\ell_m^{-1}+t_m\ell_m\right) \bm{\phi}_m.
\end{align}
\end{subequations}
Eq.~\eqref{eq:differenceequationinterface2} which has no matrix structure and is hence the same for both components of the spinor requires the spinor direction to be the same on both sides of the interface.  Moreover, it determines the magnitude ratio:
\begin{equation}
\label{eq:discretespinrelation}
\bm{\phi}_m
    =\frac{\Delta }{E - h_m + t_m \ell_m^{-1}} \bm{\phi}_w.
\end{equation}
Together with Eq.~\eqref{eq:differenceequationinterface1}, a final relation is obtained:
\begin{widetext}
\begin{equation}
\label{eq:interfacehamiltoniandiscrete}
    \left(E -\frac{\Delta^2}{E-h_m+t_m\ell_m^{-1}} - g_1 \sigma_z + g_3 \sigma_x - t\ell^{-1} \sigma_+ \right) \bm{\phi_w} = 0.
\end{equation}
\end{widetext}
Eq.~\eqref{eq:interfacehamiltoniandiscrete} is similar in form and purpose to the effective surface Green's function \eqref{eq:effectivegreen} except it is purely in spin space since the ansatze and vanishing boundary conditions took care of position dependencies. It can be interpreted as an eigenvalue problem for the matrix $g_1\sigma_z - g_3 \sigma_x + \ell^{-1}\sigma_+$ whose eigenvalues are  
\begin{equation}
     E_\Delta \equiv E - \frac{\Delta^2}{ E - h_m + t_m \ell_m^{-1}} .
\end{equation}
With this, its energy bands are twofold and defined by the implicit equation
\begin{equation}
\label{eq:implicitenergy}
    E_{\eta} - \frac{\Delta^2}{E_{\eta}-h_m+{t_m}\ell_m^{-1}} = \eta \left({g_1^2 + g_3^2 - tg_3\ell^{-1}}\right)^{\frac{1}{2}}
\end{equation}
where $\eta=\pm$ is the band index and $\ell$ ($\ell_m$) is itself a function of energy through Eq.~\eqref{eq:differenceequationWSM} [Eq.~\eqref{eq:differenceequationmetal}]: 
\begin{subequations}
    \begin{align}
        \ell_{\pm} &= Q \pm \sqrt{Q^2 - 1},  \\
        \ell_{m,\pm} &=  P \pm \sqrt{P^2 - 1}.
    \end{align}
\end{subequations}
Here, $Q = \frac{g_1^2+g_3^2+t^2-E^2}{2g_3t}$ and $P =  \frac{h_m-E}{2t_m}$.
While it may seem at first glance that the energies are symmetric in $k_x$ due to the even parity of both $\ell$ and $\ell_m$ with respect to $k_x$, one must be careful in choosing the appropriate branch $\eta$, such that the state indeed decays away from the interface. In general, the branch may vary as a function of $\mathbf{k}_{\perp}$. The infinite lattice theory Eq.~\eqref{eq:implicitenergy} is compared to the finite model in Fig. \ref{fig:discreteplots}.

\begin{figure}
    \centering
    \includegraphics[width = \columnwidth]{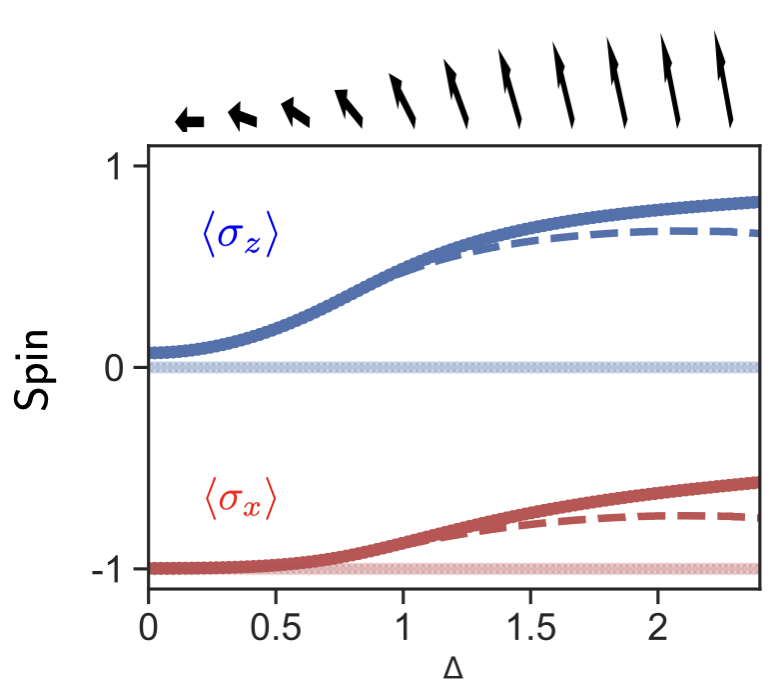}
    \caption{The chiral state's spin at the interface, fixed at the Weyl node ${k}_z = \pi / 2$ and $k_x = - 0.7$ on a lattice of size $L_y = 30$. Varying $\Delta$ cants the spin from $\sigma_x = -1$ towards $\sigma_z = +1$ (black arrows), matching the prediction of Eq.~\eqref{eq:ratioofspinsinterfacediscrete}. The solid (dashed) lines correspond to numerical (infinite lattice theory) results. In the absence of the $\cos{k_y}$ term, the spins are unchanged with tunnelling (faded red and blue points). Note that $\expval{\sigma_y}$ is always zero (see App.~\ref{sec:sigmayzero}).}
    \label{fig:spinstunnelling}
\end{figure}

Another remarkable consequence of tunnelling is spin canting. At first glance, one should not expect that non-magnetic tunnelling to a non-magnetic metal should cause the polarized spins at the interface to cant. Indeed, this plain intuition is seemingly supported by Eq.~\eqref{eq:discretespinrelation} and agrees with the finite lattice model for $\Delta \lesssim t$. To dress a more complete picture, however, we must consider the ratio of spins of the interface state which, in light of Eq.~\eqref{eq:interfacehamiltoniandiscrete}, is
\begin{equation}
\label{eq:ratioofspinsinterfacediscrete}
     \frac{\phi^{\uparrow}_w}{\phi^{\downarrow}_w} = \frac{E_{\Delta} + g_1}{-g_3}.
\end{equation}
If $\Delta > 0$ and $E < h_m + t_m \ell_m^{-1}$, we expect the interface spin to cant away from $ {\phi^{\uparrow}_w}/{\phi^{\downarrow}_w} = 0$ ($\sigma_x = -1$) towards ${\phi^{\uparrow}_w}/{\phi^{\downarrow}_w} = -1$ ($\sigma_z = +1$). Solving for ${\phi^{\uparrow}_w}/{\phi^{\downarrow}_w}$ together with Eq.~\eqref{eq:implicitenergy} yields the spins of the chiral state as they vary with tunnelling, shown in Fig. \ref{fig:spinstunnelling}. We therefore conclude that non-magnetic surface tunnelling to a non-magnetic metal can in fact induce a change in the spins of the WSM's chiral states. It is also of note that without $H_w$'s $\cos{k_y}$ term, spin canting is absent and the interface state will remain in a $\sigma_x = -1$ eigenstate independent of tunnelling. 


\section{continuum interface theory}
\label{sec:interfacetheory}

A simplified model that can capture the effect of tunneling is a linearized continuum model which is valid at long distances.  We note however that in order to satisfy the boundary conditions at the interface, it is important to keep the second derivative in the $y$-direction, as can be seen below.

\subsection{$\Delta = 0$}
We first consider the $\Delta = 0$ case, a semi-infinite WSM slab in the continuum limit. Keeping $\mathcal{O}\left(k_y^2\right)$ terms in the WSM Hamiltonian, letting $k_y = -i\partial_y$, and multiplying by $i\sigma_y$ throughout yields the differential equation ($t=1$):
\begin{equation}
\label{eq:surfacetise}
    {\partial_y \bm{\psi} } + \frac{\sigma_x}{2}{\partial^2_y \bm{\psi}} = i\sigma_y(E - t\sin{k_x}\sigma_x)\bm{\psi} + h_z \sigma_x \bm{\psi}
\end{equation}
where $h_z \equiv g_3 - t$. To hone in on the objective interface states, we take the interface to be at $y=0$ and make the ansatz ${\bm{\psi}} \propto e^{\kappa y} \bm{\phi}$, where $\bm{\phi}$ is an unspecified spinor and $\mathrm{Re}\left(\kappa\right) > 0$ such that $\bm{\psi} \to 0$ as $y \to \infty$. The differential equation \eqref{eq:surfacetise} admits four solutions for $\kappa$, of which two have a putatively positive real part:
\begin{equation}
\label{eq:decayparameters}
    \kappa_{\pm}^2 = 2\left(1 + h_z\right) \pm 2 \left[{1 + 2h_z + E^2 - \sin^2{k_x}}\right]^{\frac{1}{2}}.
\end{equation}
For a state of energy $E$, $\mathrm{Re}\left(\kappa\right) > 0$ translates to $E^2 < \sin^2{k_x} + h_z^2$, in agreement with the infinite lattice theory. 
Eq.~\eqref{eq:decayparameters} sheds light on the fact that for a given eigenvector with energy $E$ satisfying Eq.~\eqref{eq:surfacetise}, there is a distinct eigenvector with equal and opposite energy $-E$ which is also a solution. Therefore, there are two $\kappa$ values per energy. 

To determine which solution is correct, we impose the boundary condition $\bm{\psi}(0)=0$. In general, one has the superposition ${\bm{\psi}} \propto e^{\kappa_+ y} \bm{\phi}_{\kappa_+} + \alpha e^{\kappa_- y} \bm{\phi}_{\kappa_-}$. Therefore, $\alpha = -1$ and $\bm{\phi}_{\kappa_+} = \bm{\phi}_{\kappa_-}$. Equating the ratio of spins, the latter condition can be surmised as
\begin{equation}
\label{eq:ratioofspinorscontinuum}
    \frac{E + h_z - \kappa_{+}^2 / 2}{\sin{k_x}+\kappa_{+}} = \frac{E + h_z - \kappa_{-}^2 / 2}{\sin{k_x}+\kappa_{-}}.
\end{equation}
After some algebra, we recover the aforementioned chiral state of energy $E = -t\sin{k_x}$, leading to the decay parameters $\kappa_{\pm} = 1 \pm \sqrt{1 + 2 h_z}$ and spin in the negative $x$-direction:
\begin{equation}
    \bm{\psi}_{\mathrm{chiral}} \propto
    e^{ik_x+ik_z z}\left( e^{\kappa_+ y} - e^{\kappa_- y}\right)  \begin{pmatrix}
    1 \\
    -1
    \end{pmatrix}.
\end{equation}

The condition of $\mathrm{Re}\left(\kappa\right) > 0$ leads to $    \gamma < \cos{k_z}$, which is the familiar arc condition.
At the surface BZ origin $ \mathbf{k}_{\perp,0} = \left(0,0\right)$, the chiral state's decay length is on the order of a lattice length, pointing to a strongly localized state which may therefore well be described by a continuum interface theory. At the surface Weyl points $\mathbf{k}^{\pm}_{\perp,w} = (0, \pm k_w)$,
however, $\kappa_{-} = 0$ and the chiral state's decay length diverges, as expected from the absence of such surface states at the Weyl node. 

\subsection{$\Delta > 0$}
In order to get a simple analytical result, we imagine coupling the WSM to a quantum dot of energy $M$. Here, we model the metal as a flat band since it is well above the WSM and only states with the same energy are relevant. The continuum Hamiltonian reads
\begin{align}
\label{eq:interfacesubspace}
    H_{\mathrm{cont}} = \begin{pmatrix}
    \sin{k_x} \sigma_x + h_z \sigma_z - i \sigma_y \partial_y - \frac{1}{2}\sigma_z \partial^2_y  & \Delta \\
    \Delta  & M
    \end{pmatrix}.
\end{align}
Once again, we focus on solutions bound to the interface $\bm{\psi}_{w} \propto e^{\kappa y} \bm{\phi}_w$ ($\bm{\psi}_{m} \propto e^{-\kappa_m y} \bm{\phi}_m$), leading to four differential equations. The first two restrict the metal spins to be identical to the Weyl spins up to a scalar factor: 
\begin{equation}
    {\bm{\phi}_{m}} = \frac{\Delta}{E - M} {\bm{\phi}_{w}}.
\end{equation}
The remaining two equations reduce to a $2 \times 2$ matrix equation expressed in the basis of Weyl spins $\bm{\phi}_w$:
\begin{widetext}
\begin{align}
\label{eq:continuuminterfacetise}
 \left(E - \frac{\Delta^2}{E - M} - \sin{k_x} \sigma_x - h_z \sigma_z + \frac{\kappa^2}{2}\sigma_z  + i \kappa \sigma_y \right) {\bm{\phi}_w} = 0,
\end{align}
\end{widetext}
which is the continuum form of Eq.~\eqref{eq:interfacehamiltoniandiscrete}. When $\Delta = 0$ and $E \neq M$, it is not difficult to
see that the bare WSM surface chiral state is recovered. For $\Delta > 0$, the physics are identical to the $\Delta = 0$ case with the substitution $ E \to E - {\Delta^2 }/(E - M) = E_{\Delta}$ 
\footnote{In fact, the effective surface propagator Eq.~\eqref{eq:effectivesamesitepotential} exactly reduces to $-\Delta^2 / \left(E - M\right)$ when $t_m = 0$ and $\mu = -M$.}. 
For instance, the decay parameters are now
\begin{equation}
    \kappa_{\pm}^{2} a^2t = 2 \left(t + h_z\right)  \pm 2 \left({t^2 + 2 t h_z - t^2\sin^{2}{k_x} + E_{\Delta}^2}\right)^{\frac{1}{2}},
\end{equation}
where we have re-inserted the energy scale $t$ and the lattice constant $a$. 

The continuum interface theory therefore hints at a straightforward interpretation of the energy shift upon tunnelling. Indeed, seeing as the only effect of $\Delta$ was to shift the energies, the
chiral band's energy in the continuum theory is defined by $E_{\Delta} = - t\sin{k_x}$, or
\begin{equation}
\label{eq:chiralenergytunnelling}
    E = \frac{M - t\sin{k_x}}{2} - \frac{1}{2} \left[{\left(M + t\sin{k_x} \right)^2 + 4 \Delta^2}\right]^{\frac{1}{2}}.
\end{equation}

\begin{figure}
    \centering
    \includegraphics[width = \columnwidth]{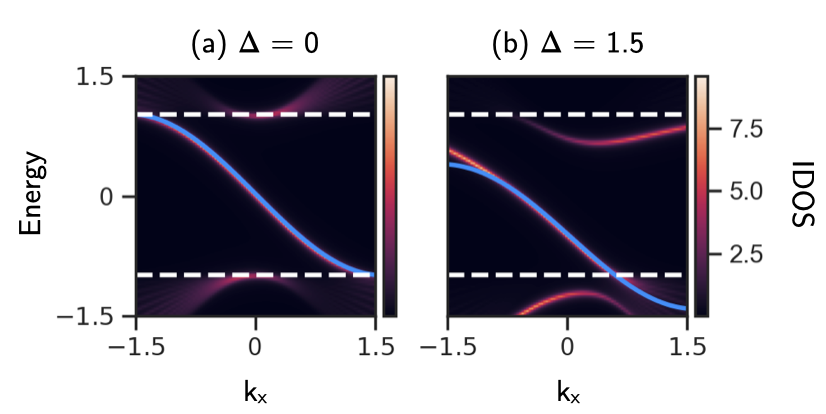}
    \caption{Interface density of states for the WSM-metal system ($L_y = 30$) at $k_z = 0$ for a tunnelling strength of (a) $\Delta = 0$ and (b) $\Delta = 1.5$. The blue line is the analytic chiral state dispersion $E_{\mathrm{chiral}}$ whereas the dashed white lines represent the bulk energy gap $E_{\mathrm{bulk}} = \pm 1$. The metal energy is $M = 4$.}
    \label{fig:gammaspectrum}
\end{figure}

In regimes where the decay lengths $\kappa^{-1}_{\pm} \sim a$, Eq.~\eqref{eq:chiralenergytunnelling} is in agreement with finite lattice simulations, as shown in Fig.~\ref{fig:gammaspectrum}. As for the chiral state's spin, it remains unchanged due to Eq.~\eqref{eq:ratioofspinorscontinuum} still being satisfied and equal to $-1$ when $E\to E_{\Delta} = -t \sin{k_x}$\footnote{In the infinite lattice theory, the replacement $$E \to E_{\Delta} = -t\sin{k_x} =-g_1$$ in Eq.\eqref{eq:ratioofspinsinterfacediscrete} also leads to a spin $\sigma_x = -1$ ($r_{\mathrm{interface}} = 0$).}. Therefore, the validity of Eq.~\eqref{eq:chiralenergytunnelling} will depend wholly on whether or not the state is in a $\sigma_x=-1$ eigenstate, and any deviations in the bandstructure must reflect a changing spin in the lattice model. Since the spins do in fact cant for $\Delta \gtrsim t$, this is the root of the continuum theory's inaccuracy in this regime. 

Another aspect captured by the continuum theory is the localization of bulk states at the interface to produce the emergent interface state, a typical feature of systems with boundary topologies \cite{andreevedgestates}. Simply put, the lowering of energy with tunnelling will give the bulk state's decay parameter a positive real part. 


\section{Transport}
\label{sec:transport}

\subsection{Along the interface}
\label{sec:transportalong}

We will now turn to the transport consequences of the previously described theory and numerics. We begin by analyzing the current along the interface, travelling in the $x$-direction. We fix $k_z$ and analyze transport in 2D, summing over all momenta at the end.

At $\Delta = 0$, the conductance at the Weyl node should vanish due to the gap closure and subsequent absence of uni-dimensional current-carrying states. For $\Delta > 0$, however, the presence of interface states near the Weyl node and the resulting spectral asymmetry in $k_x$ [Fig.~\ref{fig:discreteplots} (row 3, col. b)] suggests a jump in group velocity $\partial_{k_x} E$ across the Weyl point, leading to a nonzero conductance. 

We verify our reasoning numerically via the Landauer-Büttiker formalism, where conductance along the interface $\mathcal{G}_{\parallel}$ is defined as \cite{datta_1995}
\begin{equation}
    \mathcal{G}_{\parallel}(E) = \frac{e^2}{h} \mathrm{Tr}(G^{R} \Gamma_{l} G^A \Gamma_{r}).
\end{equation}
Here, $G^R$ is the usual retarded Green's function
\begin{equation}
\label{eq:advancedgreenfunction}
    G^R = \left({E - H - \Sigma^R}\right)^{-1}
\end{equation}
with the lead self-energy $\Sigma^R = \Sigma^R_l + \Sigma^R_r$ giving the quasiparticles a finite lifetime. The $\Gamma_l$ and $\Gamma_r$ operators describe the loss of electrons into the left and right leads, respectively: 
\begin{equation}
    \Gamma_{l(r)} = i \left(\Sigma^R_{l(r)} - \Sigma^A_{l(r)}\right) = -2\, \mathrm{Im}(\Sigma^R_{l(r)}).
\end{equation}
In the simplest case, we place two leads, one on each of the $x$-boundaries, which span the entire sample in the $y$-direction. Since the leads are (the interface is) in the plane perpendicular to $x$ ($y$), our construction forces the sample to be open in both the $x$-direction and the $y$-direction while still remaining periodic in $z$. For any $k_z$, $\Sigma^R_l$ takes the form
\begin{equation}
    \left(\Sigma^R_l\right)_{x,x';y,y'} = - \frac{i}{2\tau} \delta_{x0}\delta_{xx'} \delta_{yy'} ,
\end{equation}
where $\tau$ is the quasiparticle's lifetime. For its part, $\Sigma^R_r$ admits a similar form with $\delta_{x0}$ replaced by $\delta_{x,L_x-1}$. The tunnelling matrix $T$, while unchanged in the $y$-direction, now adopts a new diagonal sub-component in the $x$-direction:
\begin{equation}
    T_{x,x';y,y'} = {\Delta} \delta_{xx'} \delta_{y0}\delta_{y'L_y} .
\end{equation}

\begin{figure}
    \centering
    \includegraphics[width = \columnwidth]{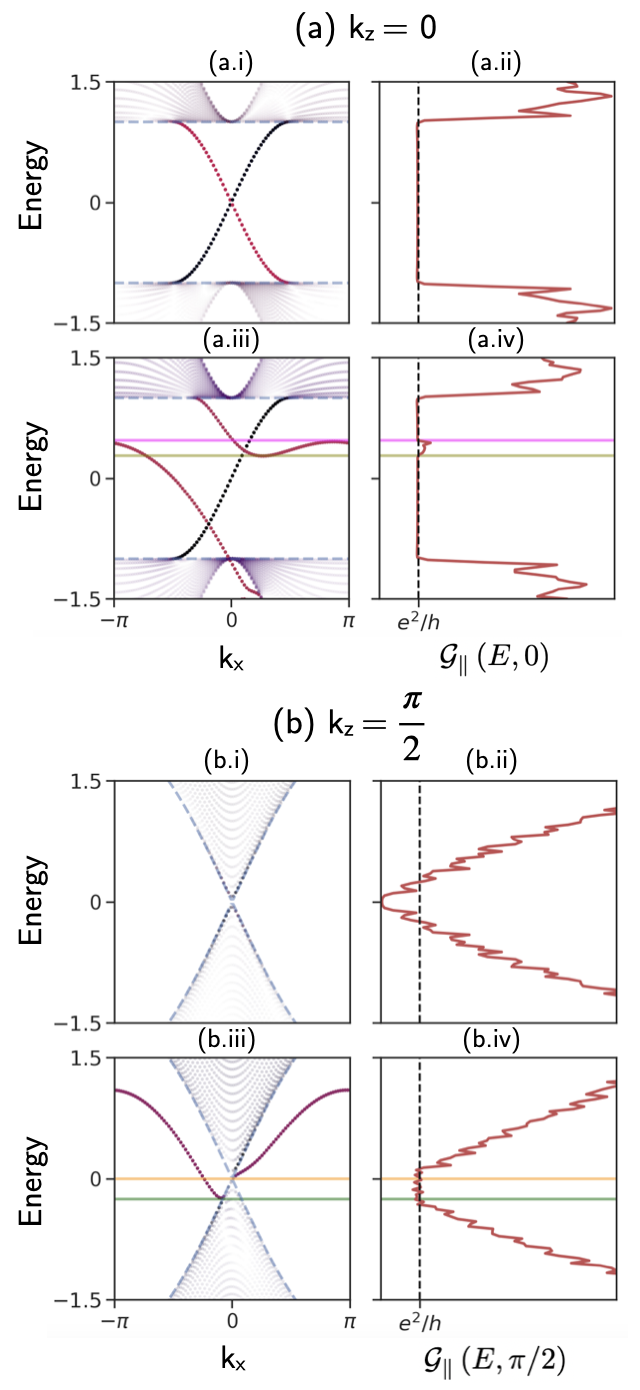}
    \caption{Conductance $\mathcal{G}_{\parallel}$ of the WSM-metal system along the interface at (a) $k_z = 0$ and (b) $k_z = \pi/2$ for $\Delta = 0$ (i, ii) and $\Delta = 2.3$ (iii, iv). To guide the physical intuition, the spectra are shown in the left panels (i, iii) and states are colored and shaded according to their $y$-position, with the relevant interface states in dark magenta and bulk states in faint colours. Energies relevant to the discussion in Sec.~\ref{sec:transportalong} are denoted by full horizontal lines.}
    \label{fig:conductancealong}
\end{figure}

For $\Delta = 0$ (Fig.~\ref{fig:conductancealong}, panels a and b, top row), the $e^2/h$ quantized conductance for $|k_z|<k_w$ can be understood in the context of the quantum anomalous Hall effect, treating each constant $k_z$ plane as a 2D quantum spin Hall insulator with one-dimensional edge states carrying $\mathcal{G}_{\parallel} = e^2/h$ \cite{bhz}. 

At $k_z = 0$ (Fig.~\ref{fig:conductancealong}a), the surface tunnelling localizes a bulk state to within the gap, allowing for both left- and right-moving carriers to produce a ``bump" in the conductance. One can reason by examining the juxtaposed spectrum. Above and below the bump energies (denoted by pink and green lines), there is only one left-moving state, whereas within it there are two left- and one right-mover. Without scattering between left and right movers, these states should contribute $2e^2/h$ to the conductance in one direction and $e^2/h$ in the other direction. On the other hand, scattering may reduce the conductance since a left and right mover can hybridize. In our case, the scattering is provided by the leads and therefore the resulting conductance is between 1 and 2 quanta of conductance.

The effect of tunnelling is perhaps most pronounced at the Weyl node (Fig.~\ref{fig:conductancealong}b). As discussed, the bulk gap closes and the subsequent absence of interface states at $\Delta=0$ leads to zero conductance at zero energy. However, with tunnelling there are now two interface states in the spectrum: the left-moving chiral state and the right-moving emergent interface state. The former will terminate at an energy $E_{\mathrm{term}}$ (green line), the intersection of $E_{\mathrm{bulk}} = \left(t^2\sin^2{k_x}+h_z^2\right)^{\frac{1}{2}}$ with Eq.~\eqref{eq:implicitenergy}.
Below $E_{\mathrm{term}}$, there are no uni-directional carriers and the conductance is unchanged. For $E_{\mathrm{term}} < E < 0$, only the chiral state is present, and there is a conductance $e^2/h$. Note the deviation from $e^2/h$ due to the small amount of bulk states present near zero energy. Above this range, both the chiral and the emergent interface state are present and move in opposite directions -- their sum is null (modulo scattering), and only bulk states contribute. 

\begin{figure}
    \centering
    \includegraphics[width=\columnwidth]{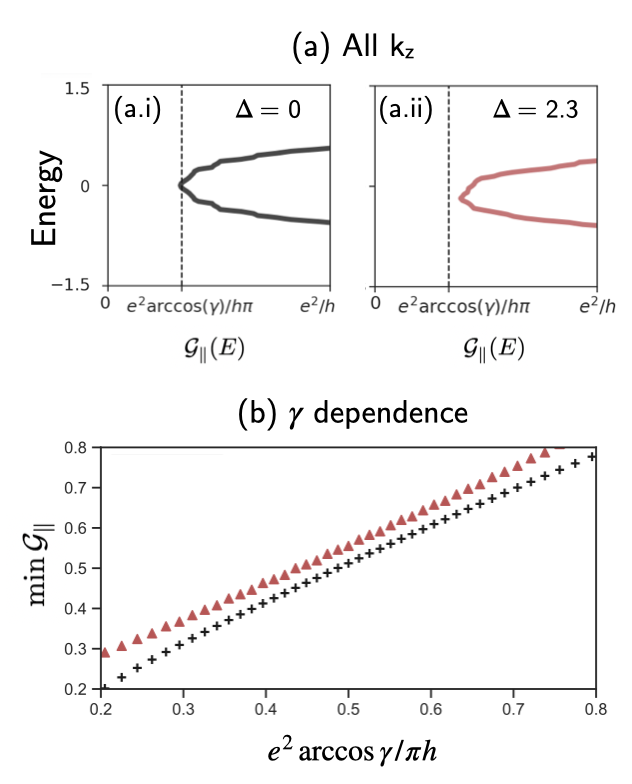}
    \caption{(a) Conductance summed over all $k_z$ for $\Delta = 0$ (i) and $\Delta=2.3$ (ii). (b) Total conductance minimum \eqref{eq:summedconductancealongmin} as a function of bare Fermi arc length. For $\Delta = 0$ (black crosses), $k^z_{\mathrm{arc}} = 2\arccos{\gamma}$ and the minimum conductance scales with the Fermi arc length (modulo scattering). For $\Delta = 2.3$ (red triangles), $k_{\mathrm{arc}}^z > 2\arccos{\gamma}$ and the conductance minimum is therefore increased relative to $\Delta=0$.}
    \label{fig:conductance_allkz}
\end{figure}

When experimentally measuring transport between leads, the measured quantity is a sum over all $k_z$ momenta. We therefore define the total conductance along the interface,
\begin{equation}
\label{eq:summedconductancealong}
\mathcal{G}_{\parallel} (E) = \frac{1}{L_z} \sum_{k_z} \mathcal{G}_{\parallel}(E,k_z).
\end{equation}
Summing over quantized conductance contributions on the $z$-projected Fermi arc $k_{\mathrm{arc}}^z $, Eq.~\eqref{eq:summedconductancealong}'s minimum is fixed (Fig.~\ref{fig:conductance_allkz}):
\begin{equation}
\label{eq:summedconductancealongmin}
    \min{\mathcal{G}_{\parallel}} = \frac{e^2}{h} \frac{k_{\mathrm{arc}}^z}{2\pi}.
\end{equation}
To probe this signature, we vary the arc length along the $k_z$-direction, as shown by Fig.\ref{fig:conductance_allkz}b. In the minimal model, this can be done by applying a strong Zeeman-like magnetic field $ b_z \mathbf{e}_{z}$ coupling to spin degrees of freedom, bringing the arc length to $k_{\mathrm{arc}}^z \to 2 \arccos{(\gamma + b_z)}$ provided $b_z$ is small enough not to change the overall topological phase and that its orbital effects may be neglected.

\subsection{Across the interface}

We complete our study with a simple analytical model for electron tunnelling across the interface. We set out to derive an expression for the conductance across the interface $\mathcal{G}_{\perp,\sigma} = d I_{\sigma} / d V$ of a particle polarized with spin $\sigma$. A detailed derivation is included in App.~\ref{sec:conductanceacross}. 

We begin by expressing the current $I_{\sigma}$ of a particle with spin $\sigma$ in terms of the retarded correlation function $U_R^{\sigma\sigma'}$ \cite{PhysRevLett.8.316,Mahan2000,ryndyk_2016}:
\begin{equation}
\label{eq:currentmain}
    I_{\sigma} =  -{2e} \, \mathrm{Im}  \sum_{\sigma'}U_R^{\sigma\sigma'}(-eV).
\end{equation}
${U}^{\sigma\sigma'}_R\left(-eV\right)$ is found by computing the Matsubara correlation function ${\mathcal{U}}^{\sigma\sigma'}(i \omega_n)$ and analytically continuing $i\omega_n \to -eV + i0^+$. At finite temperature $\beta^{-1}$, we have
\begin{equation}
\label{eq:matsubaracorrelation}
    \mathcal{U}^{\sigma\sigma'}(i\omega_n) = \frac{1}{\beta}\sum_{\mathbf{k}\mathbf{q}} |T_{\mathbf{k}\mathbf{q}} |^2 \sum_{ip} g_w^{\sigma'\sigma}(\mathbf{k},ip-i\omega_n) g_m^{\sigma\sigma'}(\mathbf{q},ip).
\end{equation}
where $\mathbf{k}$  ($\mathbf{q}$) is the momentum in the WSM (metal), $T_{\mathbf{k}\mathbf{q}}$ is the tunnelling matrix element, $\omega_n$ ($p$) is a bosonic (fermionic) Matsubara frequency, and $\bm{g}_w$ ($\bm{g}_m$) is the Matsubara Green's function for the bare WSM (metal). Since states bound to the interface will not contribute to tunnelling across of it, we may consider only bulk states. The bulk Green's functions $\bm{g}_{m,w}$ are therefore
\begin{subequations}
\begin{align}
    \bm{g}_m(\mathbf{q},ip) &= \frac{1}{ip - \xi_{m}}, \\
    \bm{g}_w(\mathbf{k},ip) &= \frac{ip + \mathcal{H}^{\mathrm{bulk}}_w}{(ip - \xi_{w})(ip + \xi_{w})},
\end{align}
\end{subequations}
with the WSM (metal) dispersion $\xi_w$ ($\xi_m$).
Setting $|T_{\mathbf{k}\mathbf{q}}|^2=\Delta^2 \delta(\mathbf{k}_{\perp}-\mathbf{q}_{\perp})$, 
we perform the Matsubara frequency summation $\sum_{ip} \left({ip - \xi}\right)^{-1} = \beta n_F \left(\xi\right)$ \cite{Mahan2000}, where $n_F$ is the fermionic distribution, by splitting the denominator into partial fractions. Using $\mathrm{Im} \, (-eV + i0^{+} - \xi)^{-1} = -\delta(-eV-\xi)$, 
Eq.~\eqref{eq:currentmain} becomes
\begin{align}
\label{eq:currentcleanmain}
    I = { 2 e \Delta^2} &\sum_{\mathbf{k}_{\perp}, k_y,q_y} \Big\{{u}_{\mathbf{k}}^2 \left[n_F(\xi_{m}) - n_F(\xi_{w})\right] \delta(-eV-\xi_-) \nonumber\\&+ {v}_{\mathbf{k}}^2\left[n_F(\xi_{m}) - n_F(-\xi_{w})\right] \delta(-eV-\xi_+)\Big\},
\end{align}
where $\xi_{\pm} = \xi_m \pm \xi_w$ and
\begin{subequations}
\label{eq:coherencefactorsmain}
\begin{align}
    {u}^2_{\mathbf{k}} &= \frac{1}{2}\left(1 + t\sin{k_x}/{\xi_w} \right),\\
    {v}^2_{\mathbf{k}} &= \frac{1}{2}\left(1 - t\sin{k_x}/{\xi_w} \right).
\end{align}
\end{subequations}
Note that we have chosen the quantization axis in the $x$-direction for simplicity. More generally, the second term in Eqs.~\eqref{eq:coherencefactorsmain} is an odd function of $k_x$, $k_z$, and $\xi_w$ and will vanish when integrated over, leaving the current spin-independent. 

To proceed, we imagine placing the metal band's Fermi level $\mu_m$ in the WSM's upper band and largely above to parabolic band minimum. At low energies,
\begin{subequations}
\begin{equation}
    \xi_w = v\left(\mathbf{k}_{\perp}^2 + k_y^2\right)^{\frac{1}{2}}
\end{equation}
and
\begin{equation}
    \xi_m = {\mu}_m + \frac{1}{m}\left(2m \tilde{\mu} - \mathbf{k}_{\perp}^2\right)^{\frac{1}{2}} q_y,
\end{equation}
\end{subequations}
where $m = 1/2t_m$ and $\tilde{\mu} = \mu + \mu_m + 6t_m$ (the lattice constant is still $a=1$).
The latter expression is found by expanding near $\xi_m$'s intercept with $\mu_m$ along $q_y$, the metal's $y$-momentum. We further consider a small positive applied voltage such that particles tunnel from the upper WSM band to the metal. Thus, only the first term of Eq.~\eqref{eq:currentcleanmain} contributes. Replacing the sums by integrals, changing variables from $k_y$ to $\xi_w$ and $q_y$ to $\xi_m$, and re-inserting $\hbar$, the current is now
\begin{align}
\label{eq:currentsimpleformmain}
       I &= \frac{e}{h} \frac{m \Delta^2}{2\pi v^2} \int_{0}^{eV} d\xi_w \int \frac{d^2\mathbf{k}_{\perp}}{(2\pi)^2} \xi_w u^2_{\mathbf{k}} \nonumber \\
        &\times  \frac{\theta(\xi_w - v|\mathbf{k}_{\perp}|)\theta(2m\tilde{\mu} - \mathbf{k}_{\perp}^2)}{\sqrt{\xi_w^2/v^2 - \mathbf{k}_{\perp}^2}\sqrt{2m\tilde{\mu} - \mathbf{k}_{\perp}^2}} .
\end{align}
Note that we have applied the low-temperature limit $n_F(\xi_w) = -\theta(\xi_w)$.
The integral over $d^2\mathbf{k}_{\perp}$ can be done analytically, yielding the conductance across the interface: 
\begin{equation}
\label{eq:conductanceacross}
    \mathcal{G}_{\perp}\left(eV\right) = \frac{e^2}{h} \frac{m \Delta^2}{\left(2\pi\right)^2 v^2} eV \log\left|\frac{\varepsilon + eV }{\varepsilon - eV }\right|.
\end{equation}
For $eV \ll \sqrt{2 m v^2 \tilde{\mu} } \equiv \varepsilon$, the leading order term is quadratic in $V$:
\begin{equation}
\label{eq:expandconductance}
    \mathcal{G}_{\perp}\left(eV\right) \approx \frac{e^2}{h} \frac{2 m \Delta^2}{\left(2\pi\right)^2 v^2 \varepsilon} \left(eV\right)^2 .
\end{equation}
Eq.~\eqref{eq:expandconductance} maintains that tunnelling measurements with featureless metals reveal the density of states at the tunnelling energy, since the three-dimensional WSM's linear dispersion corresponds to a density of states proportional to $E^2$.

\section{Conclusion and discussion}
\label{sec:conclusion}

Using both lattice and continuum frameworks, we have described the behaviour of a $\mathcal{T}$-broken WSM's interface in proximity to a non-magnetic band. When coupled to this band via non-magnetic surface tunnelling, the WSM's chiral state lowers in energy and forms, together with a previously delocalized bulk state, a noticeable spin-dependent asymmetry in the interface spectrum across the Weyl nodes. 
To model this phenomenon, we derived a infinite lattice theory of the interface and compared it to finite lattice model numerical results. 
We found that the infinite lattice theory accurately described the behaviour of the chiral state in the entire Brillouin zone (BZ), from its energy asymmetry to its spin canting at the interface. The localization of bulk states and the curving of the Fermi arc was also captured by the infinite lattice theory. 
To build intuition, we also derived a simpler continuum theory of interface states which captured the physics near $\mathbf{k}_{\perp,0}$.
Using the Landauer-Büttiker formalism, we calculated the transport of Weyl electrons travelling along the interface. Due to the asymmetry and increased Fermi arc length which allows for the presence of interface states beyond $\mathbf{k}^{\pm}_{\perp,w}$, we found a quantized increase in conductance per $k_z$ at the Weyl nodes due to tunnelling. We proposed a possible probe of this increase by relating the minimum in total conductance to the Fermi arc length.
Finally, across the interface, the conductance reproduces a simple electron tunnelling experiment, revealing the WSM's density of states. 

The results obtained herein may also be understood in the context of a pseudo-magnetic theory whereby the Weyl node separation plays the role of a magnetic gauge field \cite{grushin,grushinlorentz}. 
Alternatively, one may also view tunnelling as a finite potential well. As tunnelling broadens to link more sites on either side of the interface and broadens the well, the number of bound states increases.
Indeed, \citeauthor{guidingdirac} consider a Dirac cone (intuitively thought of as two Weyl cones of opposite chirality) under a confining potential well and find qualitatively similar interface spectra shown herein, albeit with symmetric $k_x$-spectra \cite{guidingdirac}. 



Though this toy model described the minimal case of two Weyl nodes in a magnetic WSM, these nodes always come in pairs connected by Fermi arcs. It is therefore reasonable to expect that the results obtained herein will still manifest themselves in more complicated systems with, e.g., broken inversion symmetry and a greater number of Fermi arcs. Finally, the asymmetry is resolved if one also accounts for the Hamiltonian's $\mathcal{T}$-reversed partner $\sigma_y H_w^{*}\left(-\mathbf{k}\right)\sigma_y$, instead breaking inversion symmetry.



\begin{acknowledgments}
We would like to thank C.-T. Chen,  A. Grushin, and B. Levitan for helpful discussions. LG acknowledges the hospitality of the Houches School of Physics and financial support from the NSERC CGS-M scholarship. TPB acknowledges funding from NSERC and FRQNT.
\end{acknowledgments}

\appendix

\section{The full Hamiltonian}
\label{sec:hamiltonianfullform}

Recall the Hamiltonian for the full (finite-sized) system: 
\begin{subequations}
\begin{gather}
    {H} = \sum_{\mathbf{k_{\perp}}} \sum_{y,y'=-L_y+1}^{L_y} \mathbf{f}_{\mathbf{k}_{\perp},y}^{\dagger} \mathcal{H}\left(\mathbf{k}_{\perp}\right)_{y,y'} \mathbf{f}_{\mathbf{k}_{\perp},y'}, \\
    \mathcal{H}\left(\mathbf{k}_{\perp}\right) = \begin{pmatrix}
\mathcal{H}^{\mathrm{open}}_w \left(\mathbf{k}_{\perp}\right) & T^{\dagger} \\
T & \mathcal{H}^{\mathrm{open}}_m \left(\mathbf{k}_{\perp}\right)
\end{pmatrix},
\end{gather}
\end{subequations}
where
\begin{equation}
    \mathbf{f}_{\mathbf{k}_{\perp},y} = \begin{cases} 
    \mathbf{c}_{\mathbf{k}_{\perp},y} & -L_y+1 \leq y \leq 0 \\
    \mathbf{d}_{\mathbf{k}_{\perp},y} & 1 \leq y \leq L_y
    \end{cases}
\end{equation}
with $\mathbf{c}_y = \left(c_{\mathbf{k}_{\perp},y,\uparrow}, c_{\mathbf{k}_{\perp},y,\downarrow}\right)^{\top}$. $\mathcal{H}^{\mathrm{open}}$ is the partial-in-$y$ Fourier transform of $\mathcal{H}^{\mathrm{bulk}}$. There is translational invariance in both $x$ and $z$, so each block is in general a function of $\mathbf{k}_{\perp}$. 

The bare Weyl Hamiltonian $\mathcal{H}^{\mathrm{open}}_w$ is
\begin{equation}
\label{eq:wsmhamiltonianfinite}
    \mathcal{H}^{\mathrm{open}}_w = \mathbf{C}_0 \otimes \mathbf{h}_w  + \mathbf{C}_1 \otimes \mathbf{R}_w + \mathbf{C}^{\dagger}_1 \otimes \mathbf{R}^{\dagger}_w
\end{equation}
where $\mathbf{C}_0$ is the $L_y$-sized identity and $\left(\mathbf{C}_1\right)_{y,y'} = \delta_{y+1,y'}$ is the displacement operator on the lattice.
Here,
\begin{subequations}
\begin{gather}
    \mathbf{h}_w = t_x \sin{k_x} \sigma_x + t_z \left(2 + \gamma - \cos{k_x} - \cos{k_z}\right)\sigma_z, \\
    \mathbf{R}_w = \frac{i t_y}{2} \sigma_y - \frac{t_z}{2}\sigma_z
\end{gather}
are the spin-matrices corresponding to same-site and nearest-neighbour hopping, respectively.
\end{subequations}
The bare metal Hamiltonian $\mathcal{H}^{\mathrm{open}}_m$ is written in similar form ($\sigma_0$ is the identity matrix in spin):
\begin{equation}
    \mathcal{H}^{\mathrm{open}}_m = h_m \mathbf{C}_0 \otimes \sigma_0  - t_m \mathbf{C}_1 \otimes \sigma_0 - t_m \mathbf{C}^{\dagger}_1 \otimes \sigma_0
\end{equation}
with 
\begin{equation}
    h_m = -2 t_m \left(\cos{k_x} + \cos{k_z}\right) - \mu.
\end{equation}
Finally, the tunnelling term admits the simple form $\left(T\right)_{y,y'} = \Delta \delta_{0,L_y-1}$, or
\begin{equation}
    T = \begin{pmatrix}
    0 & \dots & \Delta \\
    \vdots & \ddots & \vdots \\
    0 & \dots & 0
    \end{pmatrix},
\end{equation}
once again diagonal in spin.

\section{The interface spectrum for different band configurations}
\label{sec:varymetal}

\begin{figure}[t]
    \centering
    \includegraphics[width = \columnwidth]{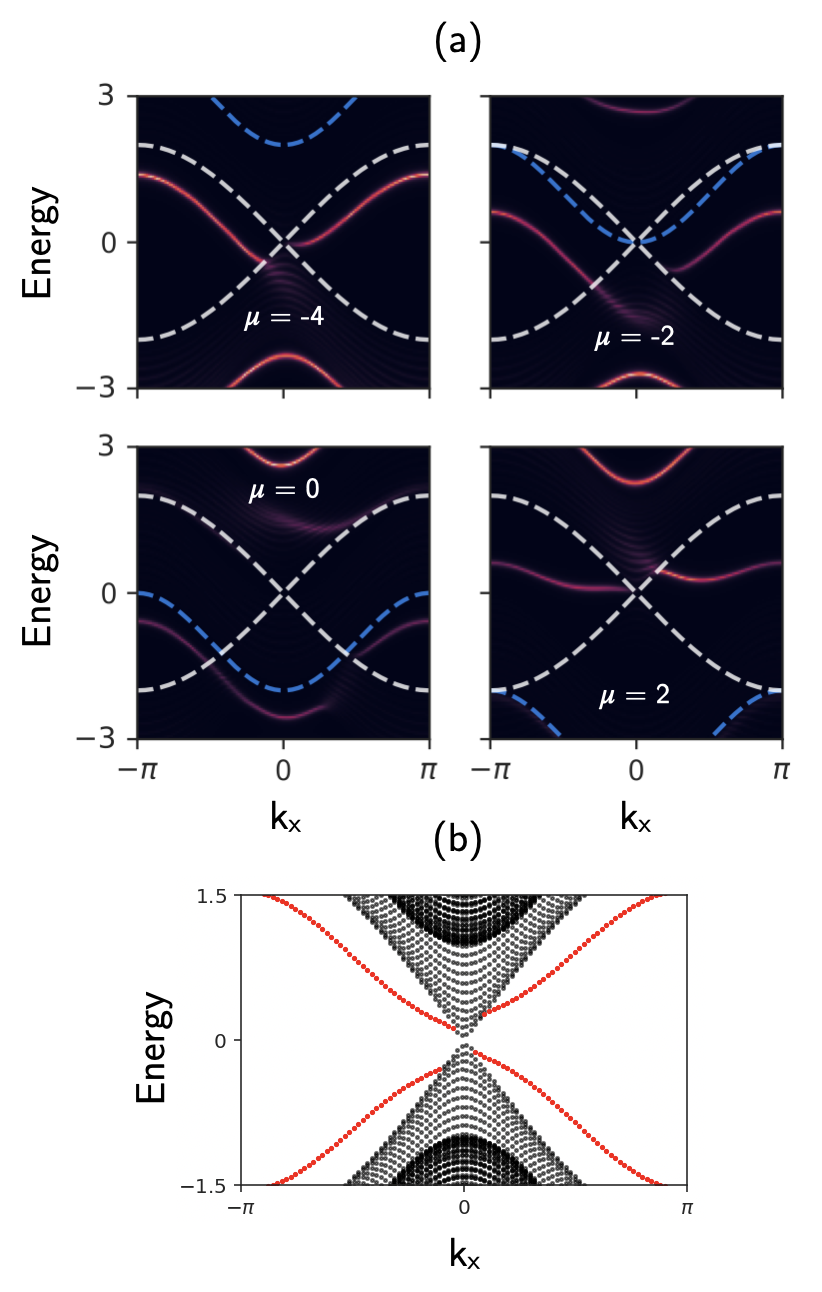}
    \caption{The energies at the Weyl node for various parabolic band configurations at $\Delta = 2.3$. (a) The interface spectral function for $t_m = 0.5$ at (from left to right, top to bottom): $\mu=-4$, $\mu=-2$, $\mu=0$, and $\mu=+2$ as defined by Eq.~\eqref{eq:metalhamiltonian}. The WSM (metal) bulk edge is shown in dashed white (blue) lines.
    (b) The bandstructure for a WSM in contact with the two-band insulator [Eq.~\eqref{eq:twobandham}]. States localized to the interface are shown in red. The insulator band gap is.}
    \label{fig:varymetal}
\end{figure}

The asymmetry at the Weyl node illustrated in Fig.~\ref{fig:discreteplots}b is also apparent for different choices of $\mu$, $t_m$, and metal bandstructure. To convince ourselves of our specific model's ubiquitous features, we display a few more metal configurations in Fig.~\ref{fig:varymetal}a. We expect that in the more realistic setup of a WSM coupled to a two-band bulk insulator, both copies will be present: one for positive and one for negative energies. This is confirmed in what follows.

One may also imagine coupling the WSM to a two-band bulk insulator, i.e. two copies of the single bulk metal band separated by a gap. Keeping each individual band non-magnetic, the Hamiltonian is now
\begin{equation}
\label{eq:twobandham}
    H = \begin{pmatrix}
    \mathcal{H}^{\mathrm{open}}_w & T^{\dagger} & T^{\dagger} \\
    T & \mathcal{H}^{\mathrm{open}}_{+} & 0 \\
    T & 0 & \mathcal{H}^{\mathrm{open}}_{-}
    \end{pmatrix}
\end{equation}
where $\mathcal{H}^{\mathrm{open}}_{\pm}$ represent the metal Hamiltonian $\mathcal{H}^{\mathrm{open}}_m$ with parameters $t_{m,\pm} = \pm t_m$, $\mu_{\pm} = \pm \mu$ and $T$ is the same as before. The resulting spectrum is shown in Fig.~\ref{fig:varymetal}b. Unsurprisingly, we recover two copies of the previously observed, single-band asymmetry: one for positive energies and one for negative energies. The asymmetry is therefore resolved if one inverts both the momentum and energy. 

Though the case of a band with zero bandwidth $t_m=0$ is not physically realistic, it still reproduces the same qualitative asymmetry. For mathematical simplification, therefore, we may set $t_m = 0$ as is done in the continuum theory and spin canting discussion. We ultimately choose to work with $\mu = -4$ and $t_m = 0.5$ and a single metal band due to the clear asymmetry across the Weyl node and separation between bulk WSM and metal dispersions.

\section{$\expval{\sigma_y} = 0$ in an open WSM}
\label{sec:sigmayzero}

In the bulk, it is clear that $\expval{\sigma_y}$ may be any value. In particular, the spin-orbit coupling in a Weyl Hamiltonian of the form $\mathbf{k}\cdot \bm{\sigma}$ will tie the $y$-momentum $k_y$ to the spin in that same direction. Upon opening our system in the $y$-direction, however, $\expval{\sigma_y} = 0$ identically throughout. Similarly, opening the system in $x$ renders $\expval{\sigma_x} = 0$.

One can see why this is the case by examining the finite-sized WSM Hamiltonian Eq.~\eqref{eq:wsmhamiltonianfinite}. Written in matrix form, the blocks are
\begin{subequations}
\begin{align}
    \mathbf{h}_w &= \begin{pmatrix}
    g_3 & g_1 \\
    g_1 & -g_3
    \end{pmatrix}, \\
    \mathbf{R}^{\dagger}_w &= \frac{1}{2} \begin{pmatrix}
    -t_z & t_y \\
    -t_y & t_z
    \end{pmatrix}
\end{align}
\end{subequations}
in spin space, where $t_{x,y,z}$ are real. Adding these blocks into the finite-sized matrix $H_w$ leads to a real and Hermitian (or, symmetric) matrix, i.e. $H_w^{\top} = H_w$. We set out to prove that one can always find real eigenstates to a real symmetric matrix, thereby rendering $\expval{\sigma_y} = 0$ identically as $\sigma_y$ is purely imaginary.

To prove this, we start by noting that the eigenvalues of a symmetric (or, more generally, Hermitian) matrix are real. Take $H_w \ket{\psi} = E \ket{\psi}$. Adding it to its complex conjugate yields
\begin{equation}
    H_w \left(\ket{\psi} + \ket{\psi}^{*}\right) = E \left(\ket{\psi} + \ket{\psi}^{*}\right).
\end{equation}
Now, if $\ket{\psi} = -\ket{\psi}^{*}$ then $\ket{\psi}$ is purely imaginary and we can therefore define $\ket{\psi} = i\ket{\varphi}$ with $\ket{\varphi}$ purely real, satisfying $H_w \ket{\varphi} = E \ket{\varphi}$. Otherwise, if $\ket{\psi}  + \ket{\psi}^{*} \neq 0$ then it is necessarily real. Therefore, one may always find a complete set of real eigenvectors to a real symmetric matrix. Since the eigenvectors are purely real and the matrix $\sigma_y$ contains only imaginary entries,
\begin{equation}
    \bra{\psi} \sigma_y \ket{\psi} = 0.
\end{equation}

Of course, one can always perform a unitary rotation in spin space such that $\sigma_x \rightarrow \sigma_y$ and $\sigma_y \rightarrow -\sigma_x$. In this case, an open system in $x$ exactly mirrors one open in $y$ before the rotation, and $\expval{\sigma_x} = 0$ likewise follows.

\section{Derivation of conductance across the interface}
\label{sec:conductanceacross}
The following procedure is similar to that performed by \citeauthor{PhysRevLett.8.316} \cite{PhysRevLett.8.316,Mahan2000,ryndyk_2016} with one important caveat: the Hamiltonian's off-diagonal elements may correlate different spins due to spin-orbit coupling. 

To begin, we posit that the current from left to right will be in direct proportion to the number of electrons leaving the Weyl side and subsequently entering the metal side. To serve our general discussion, we write the tunnelling term as
\begin{equation}
    H_T = \sum_{\mathbf{k}\mathbf{q}\sigma} T_{\mathbf{k}\mathbf{q}} c^{\dagger}_{\mathbf{k}\sigma} d_{\mathbf{q}\sigma} + \mathrm{h.c.}
\end{equation}
where $\mathbf{k}$ ($\mathbf{q}$) is the momentum on the WSM (metal) side. We further assume operators on opposite sides are independent, $\{c_i,d_j^{\dagger}\}=0$. Next, define
\begin{equation}
    N_{\sigma} = \sum_{\mathbf{k}} c_{\mathbf{k}\sigma}^{\dagger}c_{\mathbf{k}\sigma}
\end{equation}
as the number of electrons with spin $\sigma$ on the Weyl side. Since $H_w$ is number conserving, only $H_T$ fails to commute with $N_{\sigma}$:
\begin{equation}
\label{eq:heisenbergeom}
    i\hbar \dot{N}_{\sigma} = [N_{\sigma},H_T] =  \sum_{\mathbf{k}\mathbf{q}} \left( T_{\mathbf{k}\mathbf{q}} c^{\dagger}_{\mathbf{k}\sigma} d_{\mathbf{q}\sigma} - \mathrm{h.c.}\right).
\end{equation}
Though we may impose our diffusive potential $T_{\mathbf{k}\mathbf{q}} = \Delta \delta(\mathbf{k}_{\perp} - \mathbf{q}_{\perp})$ to pick out $\mathbf{k}_{\perp} = \mathbf{q}_{\perp}$, we proceed with a general tunnelling term and substitute it at the end. 

In linear response, the total current across the interface is 
\begin{equation}
    I_{\sigma}(t) = -e \expval{\dot{N}_{\sigma}(t)} = i e \int_{-\infty}^t dt^{\prime} \expval{[\dot{N}_{\sigma}(t),H_T(t')]}_0
\end{equation}
where $\expval{\cdot}_0$ represents an average over uncoupled states. To reduce the clutter of notation, define
\begin{equation}
    C_{\sigma}(t) = \sum_{\mathbf{k} \mathbf{q}} T_{\mathbf{k}\mathbf{q}} c^{\dagger}_{\mathbf{k}\sigma} (t) d_{\mathbf{q}\sigma} (t).
\end{equation}
Making the bias $\mu_m - \mu_w = eV$ explicit, it can be written as a collection of four terms:
\begin{align}
    i[\dot{N}_{\sigma}(t),H_T(t')] =& \sum_{\sigma'} \Big\{ e^{ieV(t-t')}[C_{\sigma}(t),C^{\dagger}_{\sigma'}(t')] \nonumber\\ &- e^{-ieV(t-t')}[C^{\dagger}_{\sigma}(t),C_{\sigma'}(t')]  \nonumber\\
     &+  e^{ieV(t+t')}[C_{\sigma}(t),C_{\sigma'}(t')] \nonumber\\ &- e^{-ieV(t+t')}[C^{\dagger}_{\sigma}(t),C^{\dagger}_{\sigma'}(t')] \Big\},
\end{align}
Only the first two terms are nonzero. Indeed, they represent quantum dot tunnelling, while the last two correspond to a Josephson type current, which is identically zero for the uncoupled states in the system we are considering. The first term has exactly the right form to be a retarded correlation function,
\begin{align}
        U_R^{\sigma\sigma'}(\omega) = \int_{-\infty}^{\infty} dt e^{-i\omega t} \theta(t)\expval{[C_{\sigma}(t),C^{\dagger}_{\sigma'}]}_0 
\end{align}
where we have used the fact that $\mathcal{U}_R^{\sigma\sigma'}(t)$ depends only on the time difference $t-t'$. These are precisely the kinds of terms which appear in the current calculation. We may recast $I_{\sigma}$ in terms of retarded correlation functions (re-inserting the factor of $\hbar$):
\begin{equation}
    I_{\sigma} =  -\frac{2e}{\hbar} \sum_{\sigma'} \mathrm{Im} U_R^{\sigma\sigma'}(-eV) .
\end{equation}
The Matsubara formalism provides an elegant way of finding $U_R^{\sigma\sigma'}(-eV)$ by first computing $\mathcal{U}^{\sigma\sigma'}(i \omega_n)$ and subsequently letting $i\omega_n \to -eV + i0^+$. In this framework,
\begin{align}
\label{eq:matsubaracorrelationft}
    \mathcal{U}^{\sigma\sigma'}(i\omega_n) &= - \int_0^{\beta} d\tau e^{i\omega_n \tau} \expval{T_{\tau} C_{\sigma}(\tau) C^{\dagger}_{\sigma'} }_0 \nonumber\\
    &=  \sum_{\mathbf{k}\mathbf{q}} T_{\mathbf{k}\mathbf{q}}  \int_0^{\beta} d\tau e^{i\omega_n \tau} {g}_w^{\sigma'\sigma}(\mathbf{k},-\tau) {g}_m^{\sigma\sigma'}(\mathbf{q},\tau),
\end{align}
where $\tau=it$ and $\omega_n=2\pi n /\beta$ is a bosonic Matsubara frequency because we are working with pairs of fermionic operators. $T_{\tau}$ is simply an instruction to order the operators in increasing $\tau$ starting from the right which, when combined with Wick's theorem, gives rise to the uncoupled Green's functions $\bm{g}_w(\mathbf{k},\tau)$ and $\bm{g}_w(\mathbf{q},\tau)$. 
Fourier transforming with ${g}(\tau)=\beta^{-1}\sum_{ip} e^{-ip \tau} g(ip)$ one has
\begin{equation}
    \mathcal{U}^{\sigma\sigma'}(i\omega_n) = \frac{1}{\beta}\sum_{\mathbf{k}\mathbf{q}} |T_{\mathbf{k}\mathbf{q}} |^2 \sum_{ip} g_w^{\sigma'\sigma}(\mathbf{k},ip-i\omega_n) g_m^{\sigma\sigma'}(\mathbf{q},ip).
\end{equation}

To arrive at an analytic result, we take $\bm{g}$ to be the bulk Green's function since states bound to the interface will not contribute to tunnelling across it, so we may consider only bulk states. Solving the equations of motion, we find
\begin{subequations}
\begin{align}
    \bm{g}_m(\mathbf{q},ip) &= \frac{1}{ip - \xi_{m}}, \\
    \bm{g}_w(\mathbf{k},ip) &= \frac{ip + \mathcal{H}^{\mathrm{bulk}}_w}{(ip - \xi_{w})(ip + \xi_{w})}.
\end{align}
\end{subequations}
Note that the bare WSM (metal) dispersion $\xi_w$ ($\xi_m$) depends on the momenta $\mathbf{k}$ ($\mathbf{q}$).
With 
\begin{equation}
    |T_{\mathbf{k}\mathbf{q}}|^2=\Delta^2 \delta(\mathbf{k}_{\perp}-\mathbf{q}_{\perp}),
\end{equation}
the correlation function $\mathcal{U}^{\sigma\sigma^{'}}(i\omega_n)$ can be written as a $2\times2$ matrix:
\begin{widetext}
\begin{align}
\label{eq:matsubaracorrelationmatrix}
     \bm{\mathcal{U}}(i\omega_n) =  \frac{\Delta^2}{\beta}\sum_{\mathbf{k}_{\perp},k_y,q_y} \sum_{ip} \frac{ip - i\omega_n + \mathcal{H}^{\mathrm{bulk}}_w}{(ip - \xi_{m})(ip - i\omega_n - \xi_{w})(ip - i\omega_n + \xi_{w})},
\end{align}
\end{widetext}
taking $\mathcal{H}^{\mathrm{open}}_w$ as Eq.~\eqref{eq:wsmhamiltonian}. To perform the Matsubara frequency summation, we split the denominator via partial fractions and use the relation
\begin{equation}
    \frac{1}{\beta}\sum_{ip} \frac{1}{ip - \xi} = n_F \left(\xi\right)
\end{equation}
for a fermionic frequency $ip = i \pi \left(2n + 1\right)/\beta$. Note that shifting $ip$ by a bosonic $i\omega_n$ frequency will not change the summation.
Now, upon analytic continuation, $\mathrm{Im} \, (\omega - \xi + i0^+)^{-1} = -\delta(\omega-\xi)$ relates each term to a statement of conservation of energy. We may therefore rewrite Eq.~\eqref{eq:matsubaracorrelationmatrix} as a product of occupation numbers and Dirac deltas, each weighted by spin-dependent factors. 
For an electron with spin $\sigma$, the current is therefore
\begin{widetext}
\begin{align}
\label{eq:currentclean}
    I_{\sigma} = \frac{ 2 e \Delta^2}{\hbar} \sum_{\mathbf{k}q_y}\Big\{{u}_{\mathbf{k},\sigma}^2 \left[n_F(\xi_{m}) - n_F(\xi_{w})\right] \delta(-eV-\xi_-) + {v}_{\mathbf{k},\sigma}^2\left[n_F(\xi_{m}) - n_F(-\xi_{w})\right] \delta(-eV-\xi_+)\Big\},
\end{align}
\end{widetext}
which closely resembles the superconducting case up to the modified spin-dependent coherence factors 
\begin{subequations}
\label{eq:coherencefactors}
\begin{align}
    {u}^2_{\mathbf{k},\sigma} &= \frac{1}{2}\left(1 + {\bra{\sigma} \mathcal{H}^{\mathrm{bulk}}_w \ket{\sigma}}/{\xi_w} \right),\\
    {v}^2_{\mathbf{k},\sigma} &= \frac{1}{2}\left(1 - {\bra{\sigma} \mathcal{H}^{\mathrm{bulk}}_w \ket{\sigma}}/{\xi_w} \right).
\end{align}
\end{subequations}
Here, $k_y$ ($q_y$) is momentum along $y$ in the WSM (metal), $\xi_{w}(k_x,k_y,k_z)$ [$\xi_{m}(k_x,q_y,k_z)$] is the bare WSM (metal) dispersion, $\xi_{\pm} = \xi_{m} \pm \xi_{w}$, and $n_F$ is the the usual fermionic occupation number. Eq.~\eqref{eq:currentclean} has two terms: the first (second) corresponds to tunnelling from the upper (lower) WSM band to the metal band. Each term has three parts: the Dirac delta imposes energy conservation, $n_F(\xi_{m}) - n_F(\xi_{w})$ counts the participating states available to tunnel, and $u^2_{\mathbf{k},\sigma}$/$v^2_{\mathbf{k},\sigma}$ weigh the band according to the spin. All of this is proportional to $\Delta^2$, the amplitude of a tunnelling interaction in light of the exact action Eq.~\eqref{eq:effectivegreen}.

\bibliography{apssamp}

\end{document}